%% file: arxiv.tex
\documentclass[conference]{IEEEtran}
\IEEEoverridecommandlockouts
\usepackage{cite}
\usepackage{subfigure}
\usepackage{amsmath,amssymb,amsfonts}
\usepackage{algorithmic}
\usepackage{graphicx}
\usepackage{textcomp}
\usepackage{xcolor}
\usepackage{hyperref}
\usepackage{color}
\usepackage{amsthm}
\usepackage{comment}
\usepackage{booktabs}
\usepackage{multirow}

\def\BibTeX{{\rm B\kern-.05em{\sc i\kern-.025em b}\kern-.08em
    T\kern-.1667em\lower.7ex\hbox{E}\kern-.125emX}}
\begin{document}

\ifodd 1
\newcommand{\rev}[1]{\textcolor{blue}{#1}}
\newcommand{\revw}[1]{\textcolor{red}{#1}}
\newcommand{\revg}[1]{\textcolor{cyan}{#1}}
\newcommand{\revh}[1]{#1}
\newcommand{\com}[1]{\textbf{\color{red} \left(Comment: #1\right) }}
\newcommand{\comg}[1]{\textbf{\color{blue} \left(COMMENT: #1\right)}}
\newcommand{\response}[1]{\textbf{\color{blue} \left(RESPONSE: #1\right)}}
\else
\newcommand{\rev}[1]{#1}
\newcommand{\revh}[1]{#1}
\newcommand{\revw}[1]{#1}
\newcommand{\com}[1]{}
\newcommand{\comg}[1]{}
\newcommand{\response}[1]{}
\fi

\title{FlowSpec: Continuous Pipelined Speculative Decoding for Efficient Distributed LLM Inference}

\author{
    Xing Liu$^{1*}$\thanks{$^{*}$Equal contribution $^1$Department of Computer Science and Engineering, Southern University of Science and Technology $^2$Department of Statistics and Data Science, Southern Univerisity of Science and Technology $^3$School of Computing, Montclair State University $^{4}$School of Computer Science and Engineering, Sun Yat-sen University. Corresponding author: tangm3@sustech.edu.cn} Lizhuo Luo$^{2*}$ Ming Tang$^{1}$ Chao Huang$^{3}$ Xu Chen$^{4}$
}

\maketitle

\begin{abstract}
Distributed inference serves as a promising approach to enabling the inference of large language models (LLMs) at the network edge. It distributes the inference process to multiple devices to ensure that the LLMs can fit into the device memory. Recent pipeline-based approaches have the potential to parallelize communication and computation, which helps reduce inference latency. However, the benefit diminishes when the inference request at the network edge is sparse, 
where pipeline is typically at low utilization.
To enable efficient distributed LLM inference at the edge, we propose \textbf{FlowSpec}, a pipeline-parallel tree-based speculative decoding framework. FlowSpec incorporates three key mechanisms to improve decoding efficiency: 1) score-based step-wise verification prioritizes more important draft tokens to bring earlier accepted tokens; 2) efficient draft management to prune invalid tokens while maintaining correct causal relationship during verification;
3) dynamic draft expansion strategies to supply high-quality speculative inputs.
These techniques work in concert to enhance both pipeline utilization and speculative efficiency.
We evaluate FlowSpec on a real-world testbed with other baselines.
Experimental results demonstrate that our proposed framework significantly improves inference speed across diverse models and configurations, achieving speedup ratios 1.37$\times$-1.73$\times$ compared to baselines.
Our code is publicly available at \href{https://github.com/Leosang-lx/FlowSpec#}{https://github.com/Leosang-lx/FlowSpec\#}.

\end{abstract}


\section{Introduction}\label{sec:intro}
Large language models (LLMs) have been widely adopted 
in various domains such as education, finance, and Internet of Things (IoT) \cite{minaee2024large}. However, due to their massive size, LLM inference is commonly performed at the cloud, which can introduce significant response delay due to the overload of servers during peak periods \cite{li2024llmsurvey}. Moreover, cloud-based inference often requires uploading sensitive user data, raising concerns of privacy leakage \cite{gill2025edgeaisurvey}.


In response to these limitations, deploying LLM inference at the network edge has emerged as a promising approach to achieve real-time response and enhance data privacy \cite{gill2025edgeaisurvey}, particularly for scenarios like industrial IoT and smart homes \cite{chang2021edgeaisurvey}. However, the deployment on edge devices faces two main challenges: (1) limited memory (e.g., several GBs) makes it difficult to run advanced LLMs
(e.g., LLaMA2 \cite{llama2}, Qwen2 \cite{qwen2})
with billions of parameters, and (2) restricted computational capabilities cause high inference latency. While model compression techniques like pruning \cite{movementpa} and quantization \cite{awq, smoothquant} have been explored, they often incur performance degradation and require sophisticated adaptation. A promising alternative is distributed inference, which partitions an LLM across multiple devices to enable collaborative parallel inference, which has a potential to overcome memory constraints and reduce latency.


\begin{figure}[htbp]
    \centering
    \includegraphics[width=1.0\linewidth]{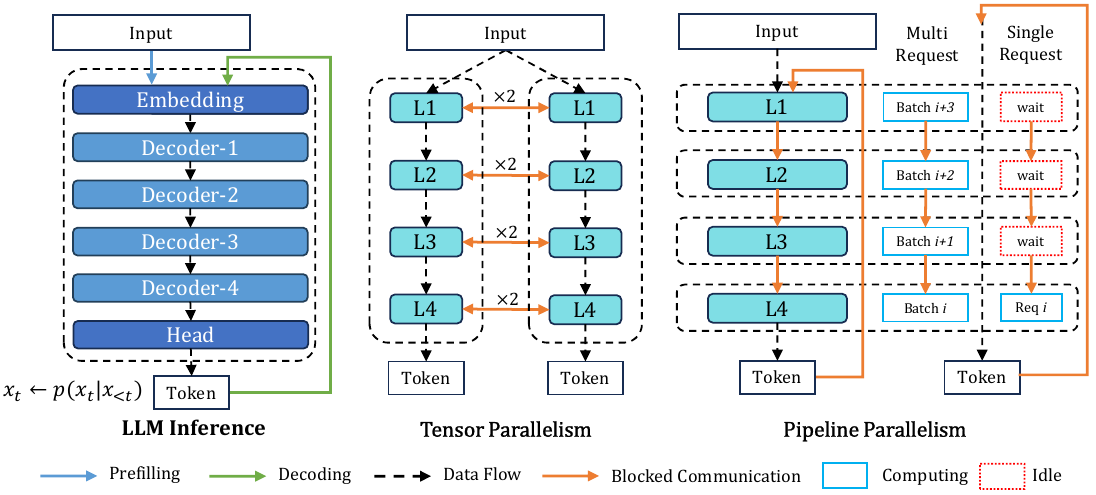}
    \caption{Left: Prefilling and decoding of LLM inference. Middle: Tensor Parallelism (TP). Right: Pipeline Parallelism with multi-/single-request. L1-L4 here represent the hidden layers.}
    \label{fig:llm_tp_pp}
    \vspace{-3mm}
\end{figure}

However, the effectiveness of distributed inference
on the edge critically depends on the underlying parallelism strategy, especially given the low-bandwidth links typical in edge scenarios.
Among common strategies, Tensor Parallelism (TP) \cite{megatron-lm, galaxy} is widely adopted in distributed inference due to its fine-grained intra-layer weight and tensor partitioning, which supports diverse inference patterns of varying batch size and sequence length.
However, TP incurs frequent synchronized all-reduce operations within each layer (as shown in Fig. \ref{fig:llm_tp_pp}), which demands high-bandwidth interconnects (e.g., NVLink, InfiniBand).
Under \textit{low-bandwidth edge connections} (e.g., less than 1000Mbps), this leads to severe communication bottlenecks that even overshadow the parallelism gains.

In contrast, Pipeline Parallelism (PP) \cite{gpipe, edgeshard} shards the model layers into sequential stages and overlaps computation and communication via micro-batching. Its coarse-grained structure reduces synchronization overhead and exhibits superior tolerance to latency and bandwidth constraints.
Hence, PP can be a natural fit for edge deployment.

However, for LLM autoregressive generation, \emph{the effectiveness of PP diminishes with sparse inference requests} (e.g., single request),
which is common at the network edge \cite{jupiter}. 
While pipelines can naturally handle multi-request inference, single autoregressive decoding cannot be parallelized across pipeline stages since a single token cannot be further split to micro-batches (as shown in Fig. \ref{fig:llm_tp_pp}),
leading to insufficient parallelism and
frequent pipeline cold starts (i.e., the period for the first input to complete all stages of an empty pipeline).
Furthermore, autoregressive decoding is inherently I/O-bound. That is, each new token requires low-compute forward propagation but large memory reads from the KV cache \cite{first_spec}.
To address this, speculative decoding (SD) \cite{first_spec} accelerates autoregressive generation by two-phase mechanism—\textbf{drafting} and \textbf{verification}.
In the \textbf{drafting phase}, a lightweight draft model autoregressively proposes candidate tokens for the subsequent $k$ positions.
Then in the \textbf{verification phase}, the base model processes all draft tokens in a single forward pass, accepting tokens sequentially as long as they satisfy the output distribution of the base LLM. In this way, SD enables generating multiple tokens in a single forward pass.
Tree-based SD \cite{specinfer, medusa, eagle} further improves token acceptance rate by organizing multiple draft sequences into a token tree, where each subsequent position (a layer of the draft tree) may host multiple candidates branching from shared prefixes.
Recent context-aware SD \cite{eagle, eagle2, GLIDEandCAPE} improves draft accuracy by enabling the draft model to leverage contextual information from the base model.
However, \emph{in conventional synchronous SD, verification must wait for full draft generation, which still limits the efficiency.} In particular, the autoregressive drafting phase remains a bottleneck, especially on edge devices with scarce computation capacity and memory bandwidth.


To this end, recent works explore asynchronous drafting and verification under pipelined architectures to accelerate SD \cite{amusd, pipeinfer, pipedec}.
By deploying the draft model and base LLM on separate devices,
verification of the current draft tokens and expanding the draft tree to generate new draft tokens
can proceed in parallel, which overlaps the latency of draft generation.
However, existing approaches still face limitations in accelerating speculative decoding with pipeline parallelism:
\begin{itemize}
    \item Existing layer-wise draft tree expansion and verification \cite{pipedec, pipeinfer} process only one layer per pipeline turn.
    This causes slow draft tree verification—without using contextual scores to prioritize tokens across layers, resulting in late token acceptance.
    \item Prior methods prune invalid drafts early but require transmission of tree structural information across stages, incurring redundant communication and tree management overhead.
    \item Existing asynchronous methods expand the draft tree from leaf nodes.
    However, in context-aware SD, this reuses stale context and fails to leverage the latest feedback from the base model. This leads to low-quality drafts, resulting in poor acceptance rates and disrupted pipeline continuity.
\end{itemize}


To address these, we propose \textbf{FlowSpec}, a pipeline-parallel speculative decoding framework for efficient distributed LLM inference with resource-constrained edge devices. Our key contributions are listed as follows:
\begin{itemize}

    \item FlowSpec introduces fast draft initialization and score-aware, step-wise verification.
    It pre-generates a deep draft tree and prioritizes high-confidence candidates in step-wise verification, enabling accepting multiple tokens in a single pipeline turn and early acceptance of deeper draft tokens to accelerate decoding.

    \item FlowSpec adopts efficient draft pruning mechanism to discard invalid draft tokens early while preserving accurate causal relationships throughout the inference process.
    Unlike existing methods,
    FlowSpec only retains the tree structural information on the draft stage to minimize extra overhead for the verification stages.

    \item FlowSpec employs new context-aware draft tree expansion strategies
    to sustain a steady supply of high-quality draft tokens.
    Instead of only expanding from the leaves, FlowSpec generates more reliable candidates from the latest context for tree expansion, improving the acceptance rate and continuity of pipeline.

    \item We evaluate FlowSpec on a real-world testbed under various settings.
    Experimental results show that FlowSpec achieves \textbf{1.37$\times$–1.73$\times$} speedups in decoding compared to state-of-the-art baselines. An ablation study further demonstrates its effectiveness.

\end{itemize}

\section{Related Works}\label{sec:p&r}

\textbf{Distributed LLM Inference on the {Edge}:}
Previous works proposed optimized TP-based approaches to mitigate communication overhead in edge scenarios. 
Voltage \cite{voltage} reduces communication overhead in TP by reordering general matrix multiplications (GEMMs) and replacing All-Reduce with All-Gather.
GALAXY \cite{galaxy} combines TP and SP for fine-grained computation-communication overlap.
But the frequent intra-layer communications for data synchronization in each decoder layer can lead to large communication and synchronization latency under the low-bandwidth connections between edges.
Other studies proposed deploying pipeline parallelism (PP) to enable collaborative LLM inference.
SARATHI \cite{chunkedprefill} proposes to split the input sequence into chunks and feed them into the pipeline progressively, enabling parallel processing during the LLM prefill phase to enhance performance.
EdgeShard \cite{edgeshard} deploys LLMs at the edge using PP, improving system throughput by load balancing and optimizing pipeline bubbles.
\emph{
However, these approaches fail to accelerate single-request LLM decoding due to the autoregressive nature.
}

\textbf{Speculative Decoding:} Speculative decoding (SD) \cite{first_spec} was proposed to accelerate LLM decoding. Built upon this framework, many works focus on tree-based SD to improve token acceptance. For example, SpecInfer \cite{specinfer} and Medusa \cite{medusa} generate multiple candidate tokens at each step, organizing them into a tree structure to enable parallel verification.
More recent works leveraged contextual information to improve the quality of draft tokens by \emph{context-aware SD}.
EAGLE \cite{eagle} predicts feature sequence of draft tokens by leveraging the hidden states from the base model and applies tree-attention when constructing the draft tree.
GLIDE and CAPE \cite{GLIDEandCAPE} utilized the KV cache of the base model to predict draft tokens and dynamically adjust the number of draft tokens.
Besides, EAGLE-2 \cite{eagle2} and OPT-Tree \cite{opt-tree} dynamically construct draft trees based on accumulated confidence scores from the draft model, enabling adaptive tree structures that improve both efficiency and draft accuracy. 
\emph{However, directly deploying these methods in distributed edge scenarios without customized computation/communication scheduling results in suboptimal performance.}

\textbf{SD with Pipeline Parallelism}. Recent works proposed extending SD to a distributed inference system to address the limitations of synchronous SD.
For example, Jupiter \cite{jupiter} partitions input sequences during prefill and enables task-level parallel SD, while it applies to only specific outline-based tasks.
AMUSD \cite{amusd} performs asynchronous drafting and verification on a two-stage pipeline, yet only supports sequence drafting.
PipeInfer \cite{pipeinfer} fills pipeline bubbles with normal decoding and draft verification, but its simple generation of multiple draft trees introduces significant redundancy across stages. Additionally, its complicated draft management degrades system efficiency.
PipeDec \cite{pipedec} considered pruning and expansion of draft trees to implement continuous SD in pipeline parallelism. However, its layer-by-layer draft tree verification significantly limits the decoding throughput, and expanding the tree only from the bottom also restricts the continuity of context-aware SD approaches in the pipeline.
In contrast, FlowSpec adopts a multi-layer expanding strategy, significantly improving the performance of speculative decoding in distributed edge scenarios.

\begin{figure*}[htbp]
    \centering
    \includegraphics[width=0.9\linewidth]{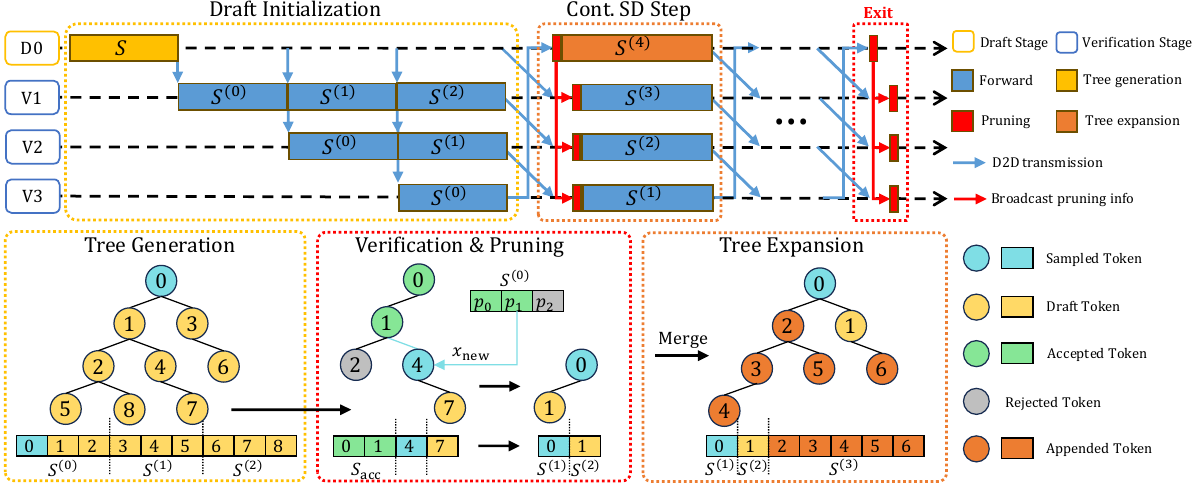}
    \vspace{-1.5mm}
    \caption{Overview of FlowSpec: An SD round includes a draft initialization step and multiple continuous SD steps. In draft initialization, segments of the initial generated draft tree are fed to V1 to fill all pipeline stages. Once the first subsequence completes all inference stages, the subsequent continuous SD steps begin. In each step, the current tree is pruned based on the accepted tokens, and then expanded to generate new draft tokens that serve as input for the next step.}
    \label{fig:system workflow}
    \vspace{-9pt}
\end{figure*}

\section{Methodology}\label{sec:methodology}

We propose FlowSpec, a pipelined tree-based speculative decoding framework for efficient distributed LLM inference targeting resource-constrained edge devices with primarily single-request scenarios.
FlowSpec prioritizes more important draft tokens for early acceptance and enables cross-layer step-wise draft verification to accelerate SD.
Besides, FlowSpec introduces an pruning mechanism
to discard invalid in real time. In addition, FlowSpec adopts novel tree expansion strategies to reduce pipeline cold starts and achieve better continuity in pipelined speculative decoding.

\subsection{FlowSpec Overview}\label{sec:overview}
Consider a distributed system with $N+1$ devices, where each device corresponds to a pipeline stage.
The first stage (with device 0) is called the \textbf{draft stage} denoted by ``D0", employs the draft model (i.e., a small, fast model that generates candidate tokens), is responsible for draft generation and accepting valid tokens based on the verification scores from the base LLM. The base LLM is partitioned into $N$ consecutive layer blocks distributed to the remaining $N$ devices, with device $n$ serving as the $n$-th \textbf{verification stage} (denoted by ``V$n$"). These stages are only responsible for executing the corresponding forward pass during the verification of draft tokens on the base LLM.

To serve an LLM generation request, the input prompt is first processed in the \textbf{prefill phase}. It is fed into the verification stages (i.e., forward pass from V1 to V$N$) to compute the initial KV cache and the first sampled token $x_\text{new}$ is generated with the base LLM. To accelerate the prefill phase for long input sequences, we adopt chunked prefill \cite{chunkedprefill}, i.e., splitting the input prompt into segments and processing them sequentially across the pipeline. 
After prefill, the system starts the \textbf{decoding phase}, which is the main focus of FlowSpec.
This phase consists of multiple SD \textbf{rounds}, an illustration of a SD round of FlowSpec is shown in Fig. \ref{fig:system workflow}. Each round starts from a draft initialization step and several following continuous SD steps:

\textbf{1) Draft Initialization Step:} D0 first generates an initial draft tree $\mathcal{T}$ with its associated draft sequence $S$ in descending order of the draft scores (see Sec. \ref{sec:score-based-draft}). $S$ is split into $N+1$ segments $S^{(0)},...,S^{(N)}$, which are then sequentially fed into the verification pipeline. As shown in Fig. \ref{fig:system workflow}, the draft segments are sequentially processed by verification stages in a pipeline-parallel manner. Once the first segment $S^{(0)}$ completes all verification stages, continuous SD step starts.
The score-based segmentation prioritizes important tokens in step-wise draft verification, enables early acceptance of draft tokens.


\textbf{2) Continuous SD Step:}
Once a draft segment $S^{(i)}$ completes all verification stages V$1$ to V$N$, its verification output (i.e., the next-token distribution from the base LLM) is sent to D0 for evaluation. With the verification result, D0 determines the accepted tokens and the new sampled token $x_\text{new}$ in this step.
If the \textbf{continuous condition} (see Sec. \ref{subsec:pruning}) is satisfied, the continuous SD steps of the current round continues and other draft segments proceed to the next verification stage. Otherwise, the current round ends and a new round is initiated.

In each continuous SD step,
and D0 executes the following two operations:
\begin{itemize}
    \item \textbf{Tree Pruning}: If D0 obtains a non-empty accepted prefix in the current continuous SD step, 
    the current tree $\mathcal{T}$ is pruned, retaining only the valid branches rooted at $x_\text{new}$. This process results in a smaller tree $\mathcal{T}_\text{pr}$. Note that all stages need to perform pruning over its local draft segment
    to discard invalids draft tokens.
    This mechanism helps to discard invalid tokens early to save redundant computation and memory overhead, while preserving the accurate causal relationship of the remaining tokens.

    \item \textbf{Tree Expansion}: At each continuous SD step, D0 extends the current draft tree in two possible ways depending on whether new tokens are accepted. If new accepted tokens exist, D0 generates a new tree $\mathcal{T}_\text{new}$ based on the updated context and merges it with $\mathcal{T}_\text{pr}$ to form the updated $\mathcal{T}$. Otherwise, D0 continues the expansion of $\mathcal{T}$ based on the extended $\mathcal{T}_\text{base}$. The newly generated draft portion then serves as the input for V1 in the next step.
    The expansion mechanism continuously provides high-quality draft tokens for the verification pipeline, brings improved acceptance rate and less pipeline cold starts.
\end{itemize}

We consider a draft initialization step accounts for $N+1$ pipeline \textbf{turns} (as shown in Fig. \ref{fig:system workflow}), i.e., one for drafting on D0 and $N$ for cold start from V1 to V$N$. A continuous SD step is considered as one pipeline \textbf{turn}, since drafting on D0 and verification of draft segments on V1 to V$N$ can be executed in parallel.

All operations of D0 associated with the draft tree are performed on CPU for efficiency, given their low parallelism, whereas the model forward computations are delegated to GPU.
In the following, unless specified, a draft sequence or segment includes its tree position IDs and attention mask.
Next, we are going to detail draft initialization in Sec. \ref{sec:score-based-draft}, and continuous SD step in Sec. \ref{subsec:pruning}, \ref{sec:tree_expansion}.

\subsection{Draft Initialization: Score-Based Draft Verification by Step}\label{sec:score-based-draft}
In the draft initialization step, D0 constructs a draft tree $\mathcal{T}$ of depth $d_0$ with $L_0$ nodes. Note that each node corresponds to a token, so we use ``node" and ``token" interchangeably. The root node is the newly sampled $x_\text{new}$ from the prefill phase or the prior SD round. The construction of $\mathcal{T}$ contains the following procedures.

\textbf{Initial Tree Generation:}
D0 sets the root node (layer 0) to be the latest $x_\text{new}$. Then, it uses the draft model to generate the initial tree $\mathcal{T}_\text{base}$.
Inspired by EAGLE-2 \cite{eagle2}, we use the cumulative confidence scores from the draft model to estimate the acceptance rate of the draft tokens.
The tree is initially rooted at $x_\text{new}$ and extended layer by layer based on the nodes with the highest cumulative scores in each layer.
The cumulative score $c_\text{cu}(n_i)$ of node $n_i$ in a draft token tree is defined as follows:
\begin{equation}\label{eq:cu_scores}
    c_\text{cu}(n_i)=\prod\limits_{n_j\in\text{path}(n_i)}c(n_j)=c_{n_i}\times c_\text{cu}(\text{parent}(n_i)),
\end{equation}
where $\text{path}(n_i)$ denotes all ancestor nodes along the path from the root node to $n_i$, and $c(n_j)$ is the confidence score of node $n_j$ from the draft model.
To construct the next layer of $\mathcal{T}_\text{base}$, the draft model takes the tokens from the previous layer as input and generates the corresponding next-token scores. Based on these scores, the top-$k$ next tokens for each input token form the new layer of $\mathcal{T}_\text{base}$. Among these, the top-$k$ nodes with the highest cumulative scores are selected to continue the expansion of the next layer. D0 expands $\mathcal{T}_\text{base}$ until it reaches layer $d_0$.
The initial depth $d_0$ serves as a hyper-parameter in FlowSpec, which will be discussed in Sec. \ref{sec:tree_params}.

\textbf{Score-Based Draft Segmentation:} Then the top-$L_0$ nodes in $\mathcal{T}_\text{base}$ with the highest cumulative scores are selected to obtain the refined draft token tree $\mathcal{T}$, with the corresponding draft token sequence $S$ sorted in descending order of the draft scores. Since a node always has a higher cumulative score than its children, the selected nodes can form a connected tree $\mathcal{T}$. Then $S$ is split into consecutive, equal-length draft segments $S^{(0)},...,S^{(N)}$ with the maximum length of $L_\text{max}$ (determined by hardware constraint). These segments are then fed into the verification pipeline one after the other.

Step-wise verification of the draft tree based on cumulative scores
offers following advantages.
First, the descending-score order prioritizes the verification of higher-scoring draft tokens, increasing the likelihood accepting more tokens in advance, thereby enabling faster and deeper expansion of the draft token tree.
Second, each draft segment consists of tokens from multiple tree layers, which allows verifying and accepting multiple tokens in a single pipeline turn (consistent with tree expansion in the following continuous SD steps). This could significantly accelerate pipelined SD compared to layer-wise verification.
In addition, since a parent node always has a higher score than its children, this ordering of the draft segments establishes a valid topological order of the tree $\mathcal{T}$, which ensures correct causal relationship between tokens in the subsequent step-wise verification (i.e., a preceding draft token must be ahead of its successors in verification).

\begin{figure*}[htbp]
    \centering
    \includegraphics[width=1.0\linewidth]{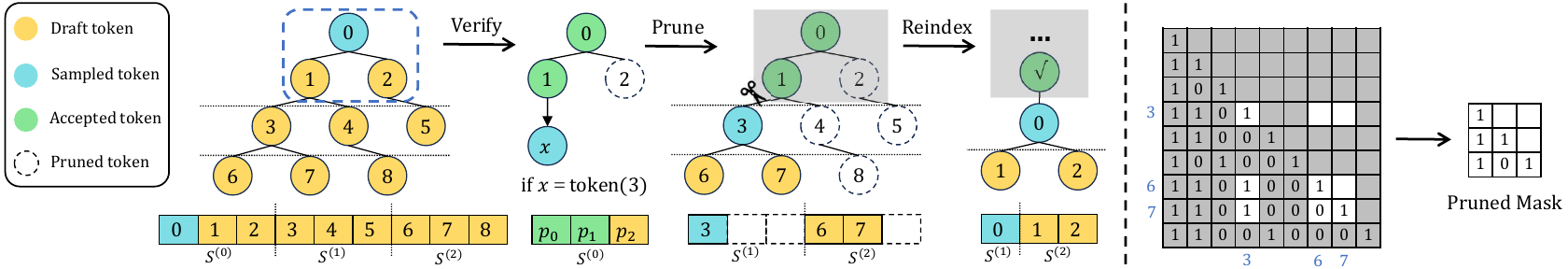}
    \vspace{-6mm}
    \caption{An example of tree pruning showing by the indices of the draft tokens. The above part is of the tree form, and the below is the sequence. In this case, $\mathcal{I}_\text{acc}=\{0,1\}$ and $\mathcal{I}_\text{draft}=\{3,6,7\}$. To prune the attention mask, only the entries whose column index and row index match the indices in $\mathcal{I}_\text{draft}$ are retained to form the updated mask.}
    \label{fig:pruning}
    \vspace{-4mm}
\end{figure*}
\subsection{Continuous SD: Tree Pruning}\label{subsec:pruning}
In each continuous SD step, once a draft segment $S^{(i)}$ completes all verification stages, D0 receives its verification output from the base LLM, and computes the output probability of $S^{(i)}$. By comparing the output distributions from the draft model and the base model, tokens that lead to an aligned distribution with the base LLM are accepted, denoted by $S_\text{acc}$. Meanwhile, an additional sampled token $x_\text{new}$ is generated based on the output probability and $S_\text{acc}$.
After that, D0 determines whether to continue the current SD round according to the \textbf{continuous condition}, which is defined as:
\vspace{-1mm}
\begin{multline}\label{continuous_condition}
    \mathtt{continuous\_condition}(\mathcal{T},S_\text{acc},x_\text{new}):=\\
    \exists n_\text{new}\in \mathcal{T}\ s.t.\ \text{path}_\mathcal{T}(n_\text{new})=\{S_\text{acc}||x_\text{new}\},
\end{multline}
where "$||$" denotes sequence concatenation. If $x_\text{new}$ satisfies the continuous condition, which means $x_\text{new}$ is already in the verification pipeline corresponding to the node $n_\text{new}\in\mathcal{T}$, thus the current SD round continues. 
\textit{Otherwise, violation of the continuous condition means that all remaining draft tokens become invalid}.
In this case, FlowSpec discards these invalid ones and their associated cache to avoid redundant overhead, exits the current SD round and initiates a new one.

The tree pruning process is illustrated in Figure \ref{fig:pruning}. To prune $\mathcal{T}$, D0 identifies all invalid branches according to $S_\text{acc}$ and $x_\text{new}$. The nodes to be retained include $n_\text{new}$ and all its descendant nodes in $\mathcal{T}$, and these nodes form the pruned tree $\mathcal{T}_\text{pr}$ with the new root node $n_\text{new}$. The corresponding pruned sequence $S_\text{pr}$ preserves the original order of the retained tokens in the original sequence $S$.

Besides D0, V1 to V$N$ also require consistent pruning to update their local draft segments. Let $\mathcal{I}_\text{acc}$ and $\mathcal{I}_\text{pr}$ denote the ordered indices of the accepted tokens and retaining tokens of $\mathcal{T}_\text{pr}$ in the sequence $S$ before pruning. D0 sends the full retaining set $\mathcal{I}_\text{retain}=\mathcal{I}_\text{acc}\cup\mathcal{I}_\text{pr}$ as the pruning information (a few hundreds of bytes) to V1-V$N$. Let $l_\text{glo}$ denote the global context offset (i.e., the total length of the input prompt and all accepted tokens), which will be updated by $l_\text{glo}=l_\text{glo}+|S_\text{acc}|$ after each pruning procedure. Then each element in $\mathcal{I}_\text{retain}$ is incremented by $l_\text{glo}$ such that the element indicates the position index of the associated token in the global sequence. For each stage, the pruning operation is applied to both the local draft segment and KV cache based on the following two index sets:

\begin{itemize}
    \item Draft segment pruning: The local positions in the current draft segment corresponding to the tokens that need to be retained in the pruned tree $\mathcal{T}_\text{pr}$,
    defined as $\mathcal{I}_\text{local}=\{i-l_\text{cache}|i\in\mathcal{I}_\text{retain},l_\text{cache}\leq i <l_\text{cache}+l_\text{s}\}$.
    Here, $l_\text{cache}$ and $l_\text{s}$ denote the length of the KV cache and the length of the draft segment at the current stage, respectively. The indices are mapped from the global positions to local draft segment positions by subtracting $l_\text{cache}$.
    \item KV cache pruning: The indices of nodes (tokens) to be retained in the KV cache are those associated with indices in set $\mathcal{I}_\text{incache}=\{i\in\mathcal{I}_\text{retain}|i<l_\text{cache}\}$. Since KV cache includes the complete context, the indices used for cache pruning correspond to global positions.
\end{itemize}

To prune a local draft segment (including its hidden states, tree position IDs and attention mask), we retain only the entries corresponding to $\mathcal{I}_\text{local}$, and the full draft sequence $S$ is accordingly pruned to be $S_\text{pr}$ consisting of the pruned segments from V1 to V$N$.
To update the KV cache, we first retain the KV cache entries from to the previous context (i.e., positions before $l_\text{glo}$). For the KV cache associated with the draft tokens (i.e., positions $\geq l_\text{glo}$), we only keep those indexed by $\mathcal{I}_\text{incache}$.
This removes redundant KV cache entries and intermediate results, ensuring both consistency and memory efficiency in subsequent decoding steps.

In this way, with keeping the structural information of draft trees only in D0, FlowSpec efficiently prunes all irregular draft segments across the distributed pipeline system with minor communication overhead, while maintaining correct causal relationship between all tokens throughout the entire lifecycle of the growing draft token tree. Early exiting the verification of invalid draft tokens further improves efficiency of speculative decoding with pipeline parallelism.

\subsection{Continuous SD: Tree Expansion}\label{sec:tree_expansion}
Unless the current SD round terminates, D0 performs tree expansion at the end of each continuous SD step.
\textbf{This process expands the current draft token tree
to sustain a continuous supply of new draft tokens for verification and hence reduces the chance of cold starts in pipeline parallelism}. There are two possible situations for tree expansion as follows:

\noindent\textbf{Context-aware Expansion.}
With new accepted tokens, D0 performs context-aware expansion immediately after pruning.
Existing draft tokens in $\mathcal{T}_\text{pr}$ are based on stale context and with less reliability for the updated context. Therefore, appending new tokens only to the bottom of $\mathcal{T}_\text{pr}$ disrupts the connection from the root node to the high-quality drafts generated from the new context in draft trees.
Instead, FlowSpec grows a new and deeper tree based on the updated context and merge it to $\mathcal{T}_\text{pr}$ as expansion, effectively improving the continuity of pipelined SD. Such continuity reduces pipeline colds, resulting in an improved efficiency of SD.


\begin{figure}
    \centering
    \includegraphics[width=1.0\linewidth]{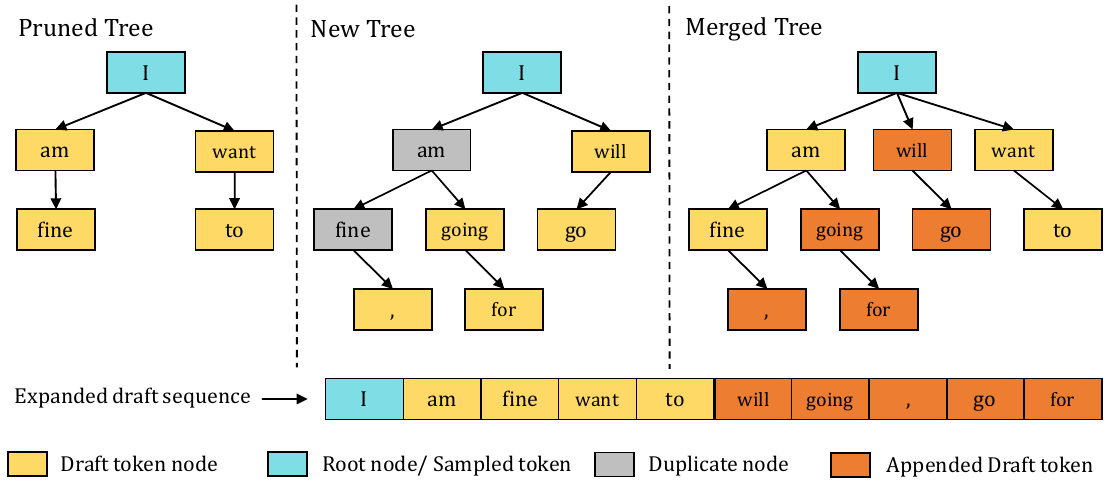}
    \vspace{-6mm}
    \caption{An example explaining merging the old pruned draft tree and the newly generated draft tree.}
    \vspace{-4mm}
    \label{fig:merge tree}
\end{figure}

In addition to the pruned tree $\mathcal{T}_\text{pr}$ rooted at $x_\text{new}$, D0 generates a new tree $\mathcal{T}_\text{new}$ of size $L_\text{exp}$ and depth $d_\text{exp}$ based on the new $S_\text{acc}$ and $x_\text{new}$.
$\mathcal{T}_\text{new}$ is constructed similarly to the initial draft tree and shares the updated $\mathcal{T}_\text{base}$ as its source. Then $\mathcal{T}_\text{pr}$ and $\mathcal{T}_\text{new}$ are merged as an updated $\mathcal{T}$ by appending only those nodes in $\mathcal{T}_\text{new}$ that are new to $\mathcal{T}_\text{pr}$.
To identify new draft tokens in $\mathcal{T}_\text{new}$, D0 checks whether their paths exist in $\mathcal{T}_\text{pr}$, using a preloaded hash table of paths $\mathcal{P}_\text{pr}=\{\text{path}_\mathcal{T_\text{pr}}(n)|n\in\mathcal{T_\text{pr}}\}$ for efficient retrieval.
Then the set of new draft tokens from $\mathcal{T}_\text{new}$ can be defined as $\mathcal{N}_\text{new}=\{n\in \mathcal{T}_\text{new}|\text{path}_{\mathcal{T}_\text{new}}(n)\notin \mathcal{P}_\text{pr}\}$. These tokens form a new draft segment $S_\text{app}$, which serves as the input for the next continuous SD step. $S_\text{app}$ is then appended $S_\text{pr}$, resulting in the updated draft sequence $S=\{S_\text{pr}||S_\text{app}\}$.
As shown in Fig. \ref{fig:merge tree}, our tree expansion strategy has the potential to expand the draft tree from root to bottom with new draft tokens based on the updated context, significantly improving the continuity of pipelined SD.

\noindent\textbf{Score-aware Expansion}.
Since pruning is applied to the remaining draft segments, some segments received by D0 may already be reduced to empty sequences.
In this case, the context is not updated.
To ensure a continuous supply of new draft tokens, D0 also expands $\mathcal{T}$ at this point.

Specifically, D0 expands the current $\mathcal{T}_\text{base}$ with additional $d_\text{se}$ layers (as mentioned in Sec. \ref{sec:score-based-draft}) to a deeper draft tree $\mathcal{T}'_\text{base}$
from which we excluded all nodes that already exist in the current $\mathcal{T}$, and select the top-$L_\text{se}$ nodes with the highest cumulative scores from the remaining ones as the expansion of $\mathcal{T}$. The new corresponding draft segment $S_\text{se}$ is also appended to the draft sequence by $S\leftarrow \{S||S_\text{se}\}$, serving as the new draft input of the verification pipeline.
As a result, we efficiently verify a larger draft tree, leading to a higher acceptance rate of draft tokens while ensuring a steady stream of new drafts for pipelined verification.


\section{Implementation Analysis}
\subsection{Tree Parameters}\label{sec:tree_params}
As introduced in Section~\ref{sec:methodology}, FlowSpec controls draft generation via three key structural parameters: tree depth ($d_0$, $d_{\text{exp}}$), tree size ($L_0$, $L_{\text{exp}}$), and tree width ($k$).

For tree size, the initial draft length $L_0$ can be set relatively large (e.g. $L > N L_{\max}$), since $T_0$ is segmented and verified incrementally across V1–V$N$. In contrast, the expansion size $L_{\text{exp}}$ must account for potential overlap with the existing tree and post-expansion pruning; empirically, setting $L_{\text{exp}}$ such that its \textit{effective} contribution approaches $L_{\max}$ helps balance verification workloads and mitigate pipeline bubbles.
While tree width ($k$) may affect the inference latency of the draft model, it empirically takes a small value ($k<L_\text{max}$), thus it makes less impacts on the system performance. We follow the empirically optimal setting of $k$ in EAGLE-2\cite{eagle2}.


As for the tree depth, it critically affects both the drafting overhead and the expected number of accepted tokens per verification step. First, the draft generation latency $t_\text{draft}$ scales linearly with depth $d$, i.e., $t_\text{draft}\approx d t_D$, where $t_D$ is the latency to execute a single forward pass of the draft model (i.e., generating one tree layer). Besides, a deeper draft tree increases the chance of accepting more tokens per verification, enhancing the expected throughput.

In the draft initialization step, a draft tree of depth $d_0$ is generated. Since this step provides the initial input for a new SD round before pipeline overlap can be established, its latency is fully exposed. Therefore, $d_0$ should balance token acceptance gains against the drafting latency, accounting for draft model accuracy, inference cost and pipeline architecture.
In each continuous SD step, while the existing draft tokens undergo verification, D0 expands the draft tree by $d_{exp}$ additional layers to produce new candidates. To maximize D0's pipeline utilization, D0 should fully exploit the available verification window without becoming the system bottleneck. Let $t_V$ denote the execution latency of each verification stage, setting $d_\text{exp}\approx t_V/t_D$ ensures balanced latency in each continuous SD step between the draft stage D0 and verification stages V1-VN, while also provides sufficiently deep exploration for subsequent speculation.
We evaluated FlowSpec’s performance under different tree depth settings in section \ref{sec:ablation_depth}, which motivates the choice of the tree depth parameters. 



\subsection{Communication}
Communication in FlowSpec consists of two parts: (i) pruning information (see Sec. \ref{subsec:pruning}) and (ii) draft segments or intermediate pipeline results. Although (i) is introduced as additional communication for cross-device draft pruning, its excessively low transmission volume—together with low-latency transmission in edge scenarios, typically ensure minimal overhead. (ii) is the inevitable transmission in pipeline systems and constitutes the main communication component in FlowSpec.

Given that (ii) dominates the communication footprint, its overhead in bandwidth-constrained edge scenarios may affect FlowSpec's practical deployment performance.
PP-based approaches, owing to their inherently lower inter-device transmission volume and opportunity for comm-comp overlap, are better suited to low-bandwidth deployment scenarios compared to TP-based methods.
Nevertheless, when bandwidth becomes excessively constrained, the resulting increase in inter-device transmission latency can no longer be effectively masked, thereby degrading the overall throughput of the pipelined inference.
First, transmission latency should be kept below the execution latency of each pipeline stage to ensure that communication overhead can be fully overlapped with computation during continuous SD steps. While this can be satisfied by excessively low bandwidth in most cases.
Further more, higher transmission latency delays the arrival of inputs at the final pipeline stage during cold-start, thereby increasing the draft initialization overhead.
However, the improved pipeline continuity migitates this issue.
We have evaluated FlowSpec with other methods under varying bandwidth in section \ref{sec:main_result}. 


\begin{table*}[t!]
  \centering
  \small
  \caption{Main results: FlowSpec vs. baselines on 6 datasets, under greedy (T=0) and stochastic (T=1) sampling (20 samples each, max length = 256). L2 and V denote LLaMA2-Chat and Vicuna-v1.3; 7B/13B indicate model sizes.}\label{tab:main_results}
   \vspace{-2mm}
  \resizebox{\textwidth}{!}{
  \small
    \begin{tabular}{cc*{6}{cc}cc}
      \toprule
        &
        & \multicolumn{2}{c}{MT-bench}
        & \multicolumn{2}{c}{HumanEval}
        & \multicolumn{2}{c}{GSM8K}
        & \multicolumn{2}{c}{Alpaca}
        & \multicolumn{2}{c}{CNN/DM}
        & \multicolumn{2}{c}{Natural Ques.}
        & \multicolumn{2}{c}{Mean} \\
      \cmidrule(lr){3-4} \cmidrule(lr){5-6} \cmidrule(lr){7-8} \cmidrule(lr){9-10} \cmidrule(lr){11-12} \cmidrule(lr){13-14} \cmidrule(lr){15-16}
      Model & Method
        & $\xi$$\uparrow$ & SR$\uparrow$
        & $\xi$$\uparrow$ & SR$\uparrow$
        & $\xi$$\uparrow$ & SR$\uparrow$
        & $\xi$$\uparrow$ & SR$\uparrow$
        & $\xi$$\uparrow$ & SR$\uparrow$
        & $\xi$$\uparrow$ & SR$\uparrow$
        & $\xi$$\uparrow$ & SR$\uparrow$ \\
      \midrule
      \multicolumn{16}{c}{Temperature = 0} \\
      \midrule
      \multirow{5}{*}{L2 7B}
      & Chunk-PP  & 6.69 & 1.00$\times$ & 7.10 & 1.00$\times$ & 6.20 & 1.00$\times$ & 6.25 & 1.00$\times$ & 4.87 & 1.00$\times$ & 5.37 & 1.00$\times$ & 6.08 & 1.00$\times$ \\
      & Mega-LM   & 2.25 & 0.34$\times$ & 2.41 & 0.34$\times$ & 2.05 & 0.33$\times$ & 2.06 & 0.33$\times$ & 1.61 & 0.34$\times$ & 1.75 & 0.33$\times$ & 2.02 & 0.33$\times$ \\
      & Galaxy    & 2.34 & 0.35$\times$ & 2.50 & 0.35$\times$ & 2.10 & 0.34$\times$ & 2.12 & 0.34$\times$ & 1.73 & 0.36$\times$ & 1.78 & 0.33$\times$ & 2.10 & 0.35$\times$ \\
      & PipeDec   & 7.33 & 1.10$\times$ & 7.49 & 1.05$\times$ & 7.25 & 1.17$\times$ & 7.19 & 1.15$\times$ & 6.17 & 1.27$\times$ & 6.78 & 1.26$\times$ & 7.03 & 1.16$\times$ \\
      & \textbf{FlowSpec} & \textbf{9.65} & \textbf{1.44}$\times$ & \textbf{10.14} & \textbf{1.43}$\times$ & \textbf{8.92} & \textbf{1.44}$\times$ & \textbf{8.84} & \textbf{1.41}$\times$ & \textbf{6.78} & \textbf{1.39}$\times$ & \textbf{7.85} & \textbf{1.46}$\times$ & \textbf{8.70} & \textbf{1.43}$\times$ \\
      \midrule
      \multirow{5}{*}{V 7B}
      & Chunk-PP  & 7.35 & 1.00$\times$ & 7.29 & 1.00$\times$ & 6.78 & 1.00$\times$ & 6.20 & 1.00$\times$ & 4.83 & 1.00$\times$ & 5.15 & 1.00$\times$ & 6.26 & 1.00$\times$ \\
      & Mega-LM   & 2.48 & 0.34$\times$ & 2.46 & 0.34$\times$ & 2.22 & 0.33$\times$ & 2.00 & 0.32$\times$ & 1.61 & 0.33$\times$ & 1.66 & 0.32$\times$ & 2.07 & 0.33$\times$ \\
      & Galaxy    & 2.56 & 0.35$\times$ & 2.54 & 0.35$\times$ & 2.26 & 0.33$\times$ & 2.05 & 0.33$\times$ & 1.72 & 0.36$\times$ & 1.69 & 0.33$\times$ & 2.14 & 0.34$\times$ \\
      & PipeDec   & 7.61 & 1.04$\times$ & 7.58 & 1.04$\times$ & 7.49 & 1.10$\times$ & 7.27 & 1.17$\times$ & 6.19 & 1.28$\times$ & 6.81 & 1.32$\times$ & 7.16 & 1.14$\times$ \\
      & \textbf{FlowSpec} & \textbf{10.56} & \textbf{1.44}$\times$ & \textbf{10.38} & \textbf{1.42}$\times$ & \textbf{9.54} & \textbf{1.41}$\times$ & \textbf{9.00} & \textbf{1.45}$\times$ & \textbf{6.87} & \textbf{1.42}$\times$ & \textbf{7.42} & \textbf{1.44}$\times$ & \textbf{8.96} & \textbf{1.43}$\times$ \\
      \midrule
      \multirow{3}{*}{L2 13B}
      & Chunk-PP  & 1.45 & 1.00$\times$ & 1.58 & 1.00$\times$ & 1.37 & 1.00$\times$ & 1.25 & 1.00$\times$ & 1.12 & 1.00$\times$ & 1.13 & 1.00$\times$ & 1.31 & 1.00$\times$ \\
      & PipeDec   & 1.64 & 1.13$\times$ & 1.69 & 1.07$\times$ & 1.63 & 1.19$\times$ & 1.60 & 1.28$\times$ & 1.47 & 1.29$\times$ & 1.55 & 1.37$\times$ & 1.59 & 1.21$\times$ \\
      & \textbf{FlowSpec} & \textbf{2.41} & \textbf{1.66}$\times$ & \textbf{2.63} & \textbf{1.66}$\times$ & \textbf{2.33} & \textbf{1.70}$\times$ & \textbf{2.11} & \textbf{1.69}$\times$ & \textbf{1.86} & \textbf{1.66}$\times$ & \textbf{1.94} & \textbf{1.72}$\times$ & \textbf{2.21} & \textbf{1.69}$\times$ \\
      \midrule
      \multirow{3}{*}{V 13B}  
      & Chunk-PP  & 1.58 & 1.00$\times$ & 1.56 & 1.00$\times$ & 1.42 & 1.00$\times$ & 1.25 & 1.00$\times$ & 1.15 & 1.00$\times$ & 1.03 & 1.00$\times$ & 1.33 & 1.00$\times$ \\
      & PipeDec   & 1.68 & 1.06$\times$ & 1.70 & 1.09$\times$ & 1.67 & 1.18$\times$ & 1.62 & 1.30$\times$ & 1.45 & 1.26$\times$ & 1.51 & 1.47$\times$ & 1.60 & 1.20$\times$ \\
      & \textbf{FlowSpec} & \textbf{2.62} & \textbf{1.66}$\times$ & \textbf{2.63} & \textbf{1.69}$\times$ & \textbf{2.38} & \textbf{1.68}$\times$ & \textbf{2.10} & \textbf{1.68}$\times$ & \textbf{1.89} & \textbf{1.64}$\times$ & \textbf{1.77} & \textbf{1.72}$\times$ & \textbf{2.23} & \textbf{1.68}$\times$ \\
      \midrule 
      \multicolumn{16}{c}{Temperature = 1} \\
      \midrule
      \multirow{5}{*}{L2 7B}
      & Chunk-PP  & 6.43 & 1.00$\times$ & 6.51 & 1.00$\times$ & 6.11 & 1.00$\times$ & 5.85 & 1.00$\times$ & 4.58 & 1.00$\times$ & 5.25 & 1.00$\times$ & 5.78 & 1.00$\times$ \\
      & Mega-LM   & 2.13 & 0.33$\times$ & 2.17 & 0.33$\times$ & 2.07 & 0.34$\times$ & 1.93 & 0.33$\times$ & 1.54 & 0.34$\times$ & 1.74 & 0.33$\times$ & 1.93 & 0.33$\times$ \\
      & Galaxy    & 2.15 & 0.33$\times$ & 2.31 & 0.35$\times$ & 2.06 & 0.34$\times$ & 1.99 & 0.34$\times$ & 1.66 & 0.36$\times$ & 1.83 & 0.35$\times$ & 2.00 & 0.35$\times$ \\
      & PipeDec   & 7.30 & 1.14$\times$ & 7.17 & 1.10$\times$ & 7.36 & 1.20$\times$ & 7.22 & 1.23$\times$ & 6.13 & 1.34$\times$ & 6.95 & 1.32$\times$ & 7.02 & 1.21$\times$ \\
      & \textbf{FlowSpec} & \textbf{9.29} & \textbf{1.44}$\times$ & \textbf{9.34} & \textbf{1.43}$\times$ & \textbf{8.66} & \textbf{1.42}$\times$ & \textbf{8.63} & \textbf{1.48}$\times$ & \textbf{6.56} & \textbf{1.43}$\times$ & \textbf{7.77} & \textbf{1.48}$\times$ & \textbf{8.38} & \textbf{1.45}$\times$ \\
      \midrule
      \multirow{5}{*}{V 7B}
      & Chunk-PP  & 6.15 & 1.00$\times$ & 6.22 & 1.00$\times$ & 6.06 & 1.00$\times$ & 5.25 & 1.00$\times$ & 4.34 & 1.00$\times$ & 4.51 & 1.00$\times$ & 5.42 & 1.00$\times$ \\
      & Mega-LM   & 2.01 & 0.33$\times$ & 2.11 & 0.34$\times$ & 1.93 & 0.32$\times$ & 1.82 & 0.35$\times$ & 1.46 & 0.34$\times$ & 1.40 & 0.31$\times$ & 1.79 & 0.33$\times$ \\
      & Galaxy    & 2.11 & 0.34$\times$ & 2.19 & 0.35$\times$ & 2.03 & 0.33$\times$ & 1.86 & 0.35$\times$ & 1.57 & 0.36$\times$ & 1.52 & 0.34$\times$ & 1.88 & 0.35$\times$ \\
      & PipeDec   & 7.31 & 1.19$\times$ & 7.38 & 1.19$\times$ & 7.12 & 1.17$\times$ & 6.87 & 1.31$\times$ & 5.94 & 1.37$\times$ & 6.56 & 1.45$\times$ & 6.86 & 1.26$\times$ \\
      & FlowSpec  & \textbf{9.11} & \textbf{1.48}$\times$ & \textbf{9.42} & \textbf{1.51}$\times$ & \textbf{8.29} & \textbf{1.37}$\times$ & \textbf{8.09} & \textbf{1.54}$\times$ & \textbf{6.12} & \textbf{1.41}$\times$ & \textbf{7.04} & \textbf{1.56}$\times$ & \textbf{8.01} & \textbf{1.48}$\times$ \\
      \midrule
      \multirow{3}{*}{L2 13B}
      & Chunk-PP  & 1.35 & 1.00$\times$ & 1.52 & 1.00$\times$ & 1.30 & 1.00$\times$ & 1.20 & 1.00$\times$ & 1.07 & 1.00$\times$ & 1.15 & 1.00$\times$ & 1.26 & 1.00$\times$ \\
      & PipeDec   & 1.65 & 1.22$\times$ & 1.70 & 1.12$\times$ & 1.63 & 1.25$\times$ & 1.59 & 1.32$\times$ & 1.47 & 1.37$\times$ & 1.58 & 1.37$\times$ & 1.60 & 1.26$\times$ \\
      & \textbf{FlowSpec} & \textbf{2.31} & \textbf{1.71}$\times$ & \textbf{2.59} & \textbf{1.70}$\times$ & \textbf{2.20} & \textbf{1.69}$\times$ & \textbf{2.04} & \textbf{1.70}$\times$ & \textbf{1.72} & \textbf{1.69}$\times$ & \textbf{1.97} & \textbf{1.71}$\times$ & \textbf{2.14} & \textbf{1.70}$\times$ \\
      \midrule
      \multirow{3}{*}{V 13B}  
      & Chunk-PP  & 1.31 & 1.00$\times$ & 1.41 & 1.00$\times$ & 1.17 & 1.00$\times$ & 1.06 & 1.00$\times$ & 1.01 & 1.00$\times$ & 0.97 & 1.00$\times$ & 1.16 & 1.00$\times$ \\
      & PipeDec   & 1.64 & 1.25$\times$ & 1.64 & 1.16$\times$ & 1.59 & 1.36$\times$ & 1.55 & 1.46$\times$ & 1.41 & 1.40$\times$ & 1.49 & 1.54$\times$ & 1.55 & 1.34$\times$ \\
      & \textbf{FlowSpec} & \textbf{2.24} & \textbf{1.71}$\times$ & \textbf{2.33} & \textbf{1.65}$\times$ & \textbf{2.02} & \textbf{1.73}$\times$ & \textbf{1.81} & \textbf{1.71}$\times$ & \textbf{1.70} & \textbf{1.68}$\times$ & \textbf{1.64} & \textbf{1.69}$\times$ & \textbf{1.96} & \textbf{1.69}$\times$ \\
      \bottomrule
    \end{tabular}
  }
  \vspace{-8pt}
\end{table*}

\section{Experiment}\label{sec:experiment}
\vspace{-2mm}
\subsection{Setups}
\noindent\textbf{Hardware.} The testbed consists of 5 \texttt{NVIDIA Jetson Orin Nano} \cite{orinnano} units, each equipped with 8GB CPU-GPU unified memory,
interconnected via a local area network with a maximum bandwidth of 1000Mbps.

\noindent\textbf{Models and Datasets.} We evaluated the performance of FlowSpec on 4 LLMs within the EAGLE-2 framework, i.e., LLaMA2-Chat 7B, 13B \cite{touvron2023llama} and Vicuna-v1.3 7B, 13B \cite{vicuna}. The weights of the draft models and base models are sourced from Hugging Face \cite{hf}. The method’s performance is validated across 6 different downstream tasks, i.e., multi‑turn dialogue, code generation, mathematical reasoning, instruction understanding, text summarization, and question answering. The associated datasets are MT‑bench \cite{mtbench}, HumanEval \cite{humaneval}, GSM8K \cite{gsm8k}, Alpaca \cite{alpaca}, CNN/Daily Mail \cite{cnn/dailtmail}, and Natural Questions \cite{natrual}, respectively.

\noindent\textbf{Baselines\footnote{Due to prohibitively low throughput (i.e., $\xi<1$) with 13B-int4 models on the testbed, we evaluate Mega-LM and Galaxy only with 7B-fp16 models.}.}
We compare FlowSpec against 4 representative methods in distributed edge LLM inference:
\begin{itemize}
    \item Megatron-LM (Mega-LM) \cite{megatron-lm}: It enables tensor parallelism (TP) for forward pass on the base model.
    \item Galaxy \cite{galaxy}: It enhances TP by splitting the feed-forward network (FFN) and overlaps with ring-based All-Reduce.
    \item Chunk-PP \cite{chunkedprefill, jupiter}: It accelerates long-sequence LLM inference on pipeline system by split the input sequence into chunks and feed into the pipeline to enable parallelism. Here Chunk-PP feeds the chunks of the draft tree into pipeline stages to enable parallel verification.
    \item PipeDec \cite{pipedec}: It enables asynchronous draft generation and verification by progressively expanding the draft tree layer-by-layer with PP.
\end{itemize}
Among the above methods, Mega-LM and Chunk-PP serve as baselines, while Galaxy and PipeDec represent state-of-the-art approaches. All adopt the context-aware EAGLE speculative decoding framework \cite{eagle,eagle2}, and no autoregressive decoding is included for comparison.

Implementation details are provided in Appendix \ref{app:implement_detail}.

\noindent\textbf{Metrics.} 
We consider the following metrics:

\begin{itemize}
    \item \textbf{Average Accepted Length per Second $\xi$:} The average number of tokens generated per second, which reflects the overall operational efficiency of the inference system.
    \item \textbf{Speedup Ratio (SR):} The  inference acceleration achieved compared to the baselines.
    
    \item \textbf{Average Accepted Length per Pipeline Turn (ALT)}: Total new tokens divides the total pipeline turns, demonstrating the decoding performance for pipelined methods.
    \item \textbf{Average Accepted Length per SD Round (ALR)}: Total new total tokens divides the total SD rounds, demonstrating the continuity of SD with pipeline parallelism.
\end{itemize}

To ensure accurate time measurements, each method undergoes the same warm‑up procedure prior to testing in order to equalize cache hit rates.


\subsection{Main Results}\label{sec:main_result}


\textbf{Decoding Performance.} Table \ref{tab:main_results} presents the results of FlowSpec and the baselines across multiple tasks on 4 LLMs. From the table, we observe that FlowSpec significantly outperforms both Chunk-PP and PipeDec on a variety of tasks. Across multiple tasks, FlowSpec achieves an average speedup of up to \textbf{1.73$\times$} compared to Chunk-PP on the mathematical reasoning task.

Under the greedy sampling (i.e., Temperature=0), the decoding performance of FlowSpec far surpasses that of PipeDec and Chunk-PP, with speedup rates up to 1.46$\times$ for 7B models and 1.72$\times$ for 13B models.
As for stochastic sampling (i.e., Temperature=1), experimental results highlights FlowSpec's remarkable advantage. FlowSpec achieves faster decoding than baselines across all six tasks, with speedup rates upto 1.56$\times$ for 7B models and 1.73$\times$ for 13B models.
In general, FlowSpec obtains approximately average speedup rates of 1.45$\times$ for 7B models and 1.70$\times$ for 13B models compared to Chunk-PP, and about 1.2-1.4$\times$ speedups compared to PipeDec.
The speedup of FlowSpec is also consistent for both greedy sampling and stochastic sampling. Owing to the uncertainty inherent in stochastic sampling, the acceleration achieved across different datasets and models exhibits appreciable variability.

\begin{table}[t]
  \centering
  \small
  \caption{ALT \& ALR. Average accepted length per pipeline turn/SD round for methods under temperature=0 (for all six datasets).}\label{tab:avg_AL}
  \resizebox{0.48\textwidth}{!}{
    \begin{tabular}{cccccc}
    \toprule
    Model & Method & ALT-7B & ALR-7B & ALT-13B & ALR-13B\\
    \midrule
    \multirow{3}{*}{L2}
    & Chunk-PP & 0.54 & 4.89 & 0.55 & 5.02\\
    & PipeDec & 0.59 & 5.68 & 0.59 & 5.76\\
    & \textbf{FlowSpec} & \textbf{1.01} & \textbf{8.67} & \textbf{1.01} & \textbf{8.93}\\
    \midrule
    \multirow{3}{*}{V}
    & Chunk-PP & 0.55 & 5.00 & 0.55 & 5.02\\
    & PipeDec & 0.60 & 5.94 & 0.60 & 5.81\\
    & \textbf{FlowSpec} & \textbf{1.02} & \textbf{9.96} & \textbf{1.04} & \textbf{9.28}\\
    \bottomrule
    \end{tabular}
  }
  \vspace{-2mm}
\end{table}

\textbf{ALT \& ALR.}
Table \ref{tab:avg_AL} shows the mean value of average accept length of each pipeline \textbf{turn} and SD \textbf{round} (explained in Sec. \ref{sec:overview}) across four datasets.
We can approximately consider that all pipeline turns have equal latency across different methods, thus ALT reflects the overall decoding performance. Results show that FlowSpec has a larger average accepted length for each pipeline turn compared to the baselines, demonstrating its effectiveness in accelerating speculative decoding. Moreover, the ALR of FlowSpec exhibits significantly improved continuity on speculative decoding with pipeline parallelism, enabling generating more tokens in a single SD round and reduced pipeline cold starts, leading to increased decoding efficiency.

While the layer-wise verification design of PipeDec limits the throughput of speculative decoding, FlowSpec unlocks the performance of state-of-the-art speculative decoding approaches by accepting more tokens in a single pipeline step. The novel tree expansion strategy also improves the continuity of context-aware speculative decoding on pipeline, which further boosts efficiency.

\begin{figure}
    \centering
    \includegraphics[width=\linewidth]{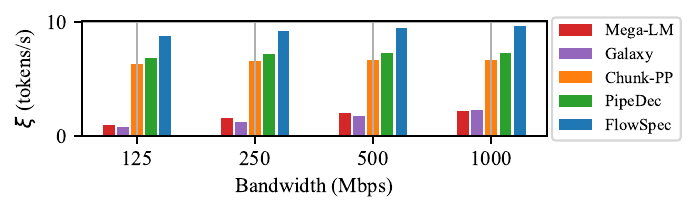}
    \vspace{-7.5mm}
    \caption{FlowSpec's $\xi$ under different bandwidths for LLaMA2-chat on MT-bench.}
    \vspace{-4mm}
    \label{fig:bw}
\end{figure}

\textbf{Performance on Varying Bandwidth}
Figure \ref{fig:bw} illustrates the performance $\xi$ of the aforementioned methods under varying inter-device bandwidth conditions. As observed, the performance of the PP-based approaches are less sensitive to bandwidth variations compared to the TP-based ones. Specifically, when the bandwidth decreases from 1000 Mbps to 125 Mbps, the PP-based methods retain over 90\% of its original performance, whereas TP-based methods suffers a substantial degradation, dropping to approximately 40\% of its original throughput. Moreover, across all bandwidth constraints, FlowSpec consistently outperforms other baselines, achieving a speedup over 1.37$\times$ compared to Chunk-PP.

\begin{table*}[t]
  \centering
    \caption{Ablation Results using the LLaMA2-Chat 7B model, showing average acceptance length per second ($\xi$) across all six datasets under two sampling settings.}
    \vspace{-2mm}
    \resizebox{1.0\textwidth}{!}{
    \begin{tabular}{c*{6}{cc}cc}
      \toprule
        & \multicolumn{2}{c}{MT-bench}
        & \multicolumn{2}{c}{HumanEval}
        & \multicolumn{2}{c}{GSM8K}
        & \multicolumn{2}{c}{Alpaca}
        & \multicolumn{2}{c}{CNN/DM}
        & \multicolumn{2}{c}{Natural Ques.}
        & \multicolumn{2}{c}{Mean} \\
        Method
        & $\xi$$\uparrow$ & SR$\uparrow$
        & $\xi$$\uparrow$ & SR$\uparrow$
        & $\xi$$\uparrow$ & SR$\uparrow$
        & $\xi$$\uparrow$ & SR$\uparrow$
        & $\xi$$\uparrow$ & SR$\uparrow$
        & $\xi$$\uparrow$ & SR$\uparrow$
        & $\xi$$\uparrow$ & SR$\uparrow$ \\
      \midrule
      \multicolumn{15}{c}{Temperature = 0} \\
      \midrule
      Chunk-PP           & 6.69 & 1.00$\times$ & 7.10 & 1.00$\times$ & 6.20 & 1.00$\times$ & 6.25 & 1.00$\times$ & 4.87 & 1.00$\times$ & 5.37 & 1.00$\times$ & 6.08 & 1.00$\times$ \\
      Pruned PP          & 8.09 & 1.20$\times$ & 8.39 & 1.18$\times$ & 7.82 & 1.26$\times$ & 7.87 & 1.25$\times$ & 6.45 & 1.32$\times$ & 7.34 & 1.36$\times$ & 7.66 & 1.25$\times$ \\
      FlowSpec w/o SBD         & 9.09 & 1.35$\times$ & 9.69 & 1.36$\times$ & 8.66 & 1.39$\times$ & 8.74 & 1.39$\times$ & 6.63 & 1.36$\times$ & 7.82 & 1.45$\times$ & 8.43 & 1.38$\times$ \\
      \textbf{FlowSpec}  & \textbf{9.65} & \textbf{1.44$\times$} & \textbf{10.14} & \textbf{1.43$\times$} & \textbf{8.92} & \textbf{1.44$\times$} & \textbf{8.84} & \textbf{1.41$\times$} & \textbf{6.78} & \textbf{1.39$\times$} & \textbf{7.85} & \textbf{1.46$\times$} & \textbf{8.70} & \textbf{1.43$\times$} \\
      \midrule
      \multicolumn{15}{c}{Temperature = 1} \\
      \midrule
      Chunk-PP           & 6.43 & 1.00$\times$ & 6.51 & 1.00$\times$ & 6.11 & 1.00$\times$ & 5.85 & 1.00$\times$ & 4.58 & 1.00$\times$ & 5.25 & 1.00$\times$ & 5.78 & 1.00$\times$ \\
      Pruned PP          & 7.79 & 1.21$\times$ & 8.00 & 1.22$\times$ & 7.80 & 1.27$\times$ & 7.42 & 1.26$\times$ & 6.16 & 1.34$\times$ & 6.89 & 1.31$\times$ & 7.34 & 1.26$\times$ \\
      FlowSpec w/o SBD         & 8.63 & 1.34$\times$ & 8.99 & 1.38$\times$ & 8.41 & 1.37$\times$ & 8.09 & 1.38$\times$ & 6.31 & 1.37$\times$ & 7.54 & 1.43$\times$ & 7.99 & 1.38$\times$ \\
      \textbf{FlowSpec}  & \textbf{9.29} & \textbf{1.44$\times$} & \textbf{9.34} & \textbf{1.43$\times$} & \textbf{8.66} & \textbf{1.42$\times$} & \textbf{8.63} & \textbf{1.48$\times$} & \textbf{6.56} & \textbf{1.43$\times$} & \textbf{7.77} & \textbf{1.48$\times$} & \textbf{8.38} & \textbf{1.45$\times$} \\
      \bottomrule
    \end{tabular}
  }
  \vspace{-2mm}
  \label{tab:ablation}
\end{table*}

\begin{figure}
    \centering
    \subfigure[L2-7B]{
        \includegraphics[width=0.48\linewidth]{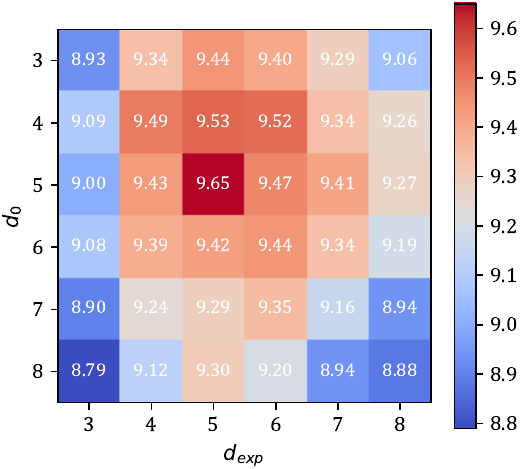}
        \hspace{-4mm}
    }
    \subfigure[V-7B]{
        \includegraphics[width=0.48\linewidth]{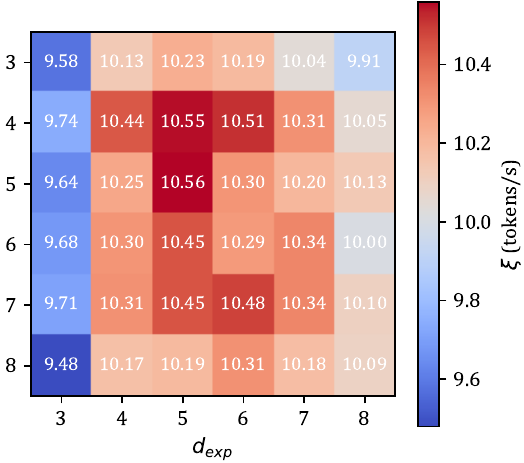}
    }
    \vspace{-3mm}
    \caption{FlowSpec's $\xi$ with varying tree depth $d_0$ and $d_{exp}$ on MT-bench.}
    \label{fig:perf_and_depths}
    \vspace{-4mm}
\end{figure}

\subsection{Ablation Study}
\subsubsection{Tree Depth Evaluation}\label{sec:ablation_depth}

As analyzed in Section \ref{sec:tree_params}, the draft tree depth may serve as a key trade-off parameter between computational cost of the draft stage and overall generation efficiency.
For deploying the EAGLE \cite{eagle} draft model and the 7B base model on our testbed, the inference latency $t_d\approx14\text{ms}$ and $t_v\approx80\text{ms}$.
Based on this latency profile, we evaluate FlowSpec across varying tree depths $d_0$ and $d_{exp}$ using the LLaMA-2-Chat model on the MT-Bench benchmark.
Results are shown in Figure \ref{fig:perf_and_depths}.
FlowSpec achieves optimal performance when the initial tree depth $d_0$ and extension depth $d_{exp}$ are both set to 5 (around $t_d/t_v$). Under this configuration, the draft stage D0 effectively saturates the time budget of each pipeline iteration closely, producing sufficiently deep draft trees and fully exploiting the draft model’s capacity. Crucially, this depth aligns the computation load of D0 with that of the subsequent verification stages (V1–V$N$), ensuring balanced stage latencies and thereby minimizing pipeline bubbles and improving overall throughput.

\subsubsection{Module Evaluation}
Table \ref{tab:ablation} presents the results for different module configurations of FlowSpec on multiple tasks using LLaMA2-Chat 7B under two sampling conditions (Temperature = 0 or 1). \textit{Chunk-PP} \cite{chunkedprefill} follows the same setting as above. \textit{Pruned Pipeline Parallelism (Pruned PP)} extends Chunk-PP by leveraging Tree Pruning.
\textit{FlowSpec w/o Score-Based Draft (FlowSpec w/o SBD)} 
omits Score-Based Draft Segmentation while applying Tree Expansion on top of Pruned PP. The different combinations of components confirm the effectiveness of each module.


\noindent{\textbf{Score-Based Draft Segmentation.}} Comparing FlowSpec to FlowSpec w/o SBD in Table \ref{tab:ablation} directly highlights the benefit of the Score-Based component. By sorting draft tokens according to contextual scores, Score-Based Draft Segmentation accepts more draft tokens early to accelerate decoding.
As a result, FlowSpec achieves an average throughput of 9.65 tokens/s and 9.29 tokens/s, both outperforming the method without Score-Based Draft Segmentation.

\noindent{\textbf{Tree Pruning.}} The difference between Chunk-PP and Pruned PP shows the effectiveness of Tree Pruning. Enabling pruning provides a significant speedup over Chunk-PP, indicating a clear improvement in performance. Removing redundant computation and communication leads to improved speed and efficiency.

\noindent{\textbf{Tree Expansion.}} 
Table \ref{tab:ablation} demonstrates substantial speedups of Tree Expansion. Notably, on the multi-turn dialogue task, FlowSpec w/o SBD achieves throughputs of 9.09 and 8.63 tokens per second at temperature settings of 0 and 1, far surpassing Pruned PP. Similar performance gains are observed across other datasets. This indicates that 
Tree Expansion improves continuity and reduces PP cold starts, leading to significant performance enhancement.

\section{Conclusion}\label{sec:conclusion}
In this work, we proposed FlowSpec, a pipeline-parallel speculative decoding framework to accelerate sparse-request distributed LLM inference with resource-constrained edge devices.
We develop three main techniques to achieve efficient continuous speculative decoding: (i) score-based step-wise verification to enable cross-layer verification and early token acceptance. (ii) efficient draft management including tree pruning and early stop to eliminate redundant computation and memory overhead. (iii) real-time tree expansion to ensure a continuous flow of high-quality draft tokens, maintaining decoding continuity. 
We evaluate FlowSpec on a real-world testbed.
Experimental results demonstrate the superiority of our method over the state-of-the-art baselines.

\bibliographystyle{IEEEtran}
\bibliography{citation}
\include{appendix}

\end{document}

%% file: appendix.tex
\appendix
\setcounter{secnumdepth}{2}

\section{Generation and Parallelism of LLMs}\label{app:llminf}

\subsection{LLM Inference and Generation}

\textbf{Decoder-only LLM Architecture:} Modern LLMs adopt a decoder-only architecture, composed of a token embedding layer followed by multiple stacked transformer decoder layers. Each decoder layer integrates self-attention mechanisms, feed-forward networks (FFNs) and residual connections, enabling LLMs to capture complex contextual dependencies. The final output of the last decoder layer is passed through a classification head to produce probability distributions for the next token. An example is illustrated in the left part of Fig \ref{fig:llm_tp_pp}.

\textbf{Generation Phases of LLMs:} The sequence generation of LLMs relies on the key-value caching (KV cache) mechanism, which divides the generation process into two phases: \textbf{prefilling} and \textbf{decoding}.
In the \textbf{prefilling phase}, the LLM processes the entire input prompt in a single forward pass to generate the probability of the first new token and save the KV cache for all initial input tokens. This phase is characterized by high computational intensity and is typically a computational bottleneck. Then in the \textbf{decoding phase}, the LLM autoregressively generates the next new token based on the previous one. Each step involves a low-FLOPs forward propagation and requires frequent access to the KV cache and model parameters, making this phase inherently I/O-bound.


\subsection{Parallelism Strategies}
The growth rate of the model parameter scales far outpaces that of hardware performance improvement, resulting in an increasing demand for distributed LLM deployment. To scale LLM inference efficiently, various parallelism strategies have been proposed. Besides TP and PP mentioned previously, other 

\textbf{Data parallelism} (DP) distributes distinct inference requests across multiple complete model replicas to enable parallel computation, typically employed to enhance the throughput of training and inference on high-performance computing (HPC) systems. However, DP becomes infeasible in edge scenarios with constrained memory and sparse requests.


\textbf{Sequence parallelism} (SP) partitions the input along the sequence dimension to enable parallel layer execution. Similar to TP, SP also requires synchronization (e.g., All-Gather) within each layer, but critically, it necessitates a complete model replica on each device. This conflicts with the memory constraints and low-bandwidth nature of typical edge environments, limiting its practicality.


In summary, \textbf{only the strategies with model partitioning} such as TP and PP can address the memory constraints of edge devices. However, they face key challenges in LLM decoding under edge scenarios. TP suffers from high communication overhead due to frequent synchronization under limited bandwidth, while PP typically struggles with low concurrency and resource utilization when dealing with sparse inference requests.
\subsection{Implementation Details}\label{app:implement_detail}
We build the testbed using \texttt{NVIDIA JetPack 5.1.2} (\texttt{L4T R35.4.1}) with \texttt{CUDA 11.4} support. We employ the ARM-aarch64-specific \texttt{PyTorch 1.11.0+cu114} \cite{torch}, installed via NVIDIA’s official Python 3.8 wheel.
Our core codebase extends the original EAGLE‑2 implementation. 

For each \texttt{NVIDIA Jetson Orin Nano} unit, its actual available GPU memory is $\sim5$GB. To deploy the 13B parameter model on the rest 4 devices, we employed W4A16 quantization (4-bit storage while preserving 16-bit computation precision) via the \texttt{bitsandbytes} library integrated in HuggingFace \texttt{transformers}.
The primary cause of the significantly lower decoding performance on the 13B models is the additional dequantization overhead.

All evaluated approaches leverage the advanced EAGLE speculative decoding framework. The draft model is deployed on a single separate device, with the base LLM partitioned across the rest deices (via TP or PP).
In the main experiments, considering that EAGLE-2 typically achieves optimal performance at a draft tree depth of 5, and our pipeline configuration, we set $L_0 = 80$, $d_{0} = 5$, and $L_{\max} = 16$ for FlowSpec and all baseline methods.
For FlowSpec, we further configure $L_{exp} = 64$ and $d_{\exp}=5$
where $d_{\exp}=5$ empirically balance the latency between the single-step draft stage and verification stages in pipeline (as stated in section \ref{sec:tree_params} and \ref{sec:ablation_depth}).
Since expanded tokens are pruned prior to transmission, the 64 newly generated tokens remain small after pruning, thereby achieving high efficiency—comparable to that of initial segments consisting of 16 tokens. For PipeDec, we select 16 tokens per layer, which aligns with its original settings. 

All other experimental settings mirror those of the main experiments.

\subsection{Limitations}
Despite the effectiveness of FlowSpec, some limitations still exist:

\textbf{Draft model dependency}: FlowSpec is designed to implement existing advanced context-aware speculative decoding frameworks with distributed edge devices more efficiently. This also implies that the overall performance is inherently bounded by the quality and efficiency of the draft models.

\textbf{Limited pipeline scalability}: scaling the pipeline with significantly more verification stages may caused degraded performance. 
First, more pipeline stages causes linearly increased transmission overhead within the inference process of each micro-batch, which is an inherent limitation of PP-based methods.
To fit more large-scale deployment scenarios, combining other model parallelism techniques can help control the number of pipeline stages while scaling efficiently.

%% file: citation.bib
@online{hf,
  title  = {``{Hugging Face}"},
  url    = {https://huggingface.co/models},
  urldate = {2025-12-07}
}

@online{orinnano,
  title  = {``{NVIDIA Jetson Orin Nano}"},
  url    = {https://www.nvidia.com/en-us/autonomous-machines/embedded-systems/jetson-orin/nano-super-developer-kit/},
  urldate = {2025-12-07}
}

@inproceedings{li2024llmsurvey,
  title={Llm inference serving: Survey of recent advances and opportunities},
  author={Li, Baolin and Jiang, Yankai and Gadepally, Vijay and Tiwari, Devesh},
  booktitle={2024 IEEE High Performance Extreme Computing Conference (HPEC)},
  pages={1--8},
  year={2024},
  organization={IEEE}
}

@article{minaee2024large,
  title={Large language models: A survey},
  author={Minaee, Shervin and Mikolov, Tomas and Nikzad, Narjes and Chenaghlu, Meysam and Socher, Richard and Amatriain, Xavier and Gao, Jianfeng},
  journal={arXiv preprint arXiv:2402.06196},
  year={2024}
}

@article{amusd,
  title={AMUSD: Asynchronous Multi-Device Speculative Decoding for LLM Acceleration},
  author={McDanel, Bradley},
  journal={arXiv preprint arXiv:2410.17375},
  year={2024}
}

@inproceedings{GLIDEandCAPE,
author = {Du, Cunxiao and Jiang, Jing and Yuanchen, Xu and Wu, Jiawei and Yu, Sicheng and Li, Yongqi and Li, Shenggui and Xu, Kai and Nie, Liqiang and Tu, Zhaopeng and You, Yang},
title = {GLIDE with a CAPE: a low-hassle method to accelerate speculative decoding},
year = {2024},
publisher = {JMLR.org},
booktitle = {Proceedings of the 41st International Conference on Machine Learning},
articleno = {465},
numpages = {17},
location = {Vienna, Austria},
series = {ICML'24}
}

@misc{vicuna,
    title = {Vicuna: An Open-Source Chatbot Impressing GPT-4 with 90\%* ChatGPT Quality},
    url = {https://lmsys.org/blog/2023-03-30-vicuna/},
    author = {Chiang, Wei-Lin and Li, Zhuohan and Lin, Zi and Sheng, Ying and Wu, Zhanghao and Zhang, Hao and Zheng, Lianmin and Zhuang, Siyuan and Zhuang, Yonghao and Gonzalez, Joseph E. and Stoica, Ion and Xing, Eric P.},
    month = {March},
    year = {2023}
}

@article{chunkedprefill,
  title={Sarathi: Efficient llm inference by piggybacking decodes with chunked prefills},
  author={Agrawal, Amey and Panwar, Ashish and Mohan, Jayashree and Kwatra, Nipun and Gulavani, Bhargav S and Ramjee, Ramachandran},
  journal={arXiv preprint arXiv:2308.16369},
  year={2023}
}

@article{gpipe,
  title={Gpipe: Efficient training of giant neural networks using pipeline parallelism},
  author={Huang, Yanping and Cheng, Youlong and Bapna, Ankur and Firat, Orhan and Chen, Dehao and Chen, Mia and Lee, HyoukJoong and Ngiam, Jiquan and Le, Quoc V and Wu, Yonghui and others},
  journal={Advances in neural information processing systems},
  volume={32},
  year={2019}
}

@article{megatron-lm,
  title={Megatron-lm: Training multi-billion parameter language models using model parallelism},
  author={Shoeybi, Mohammad and Patwary, Mostofa and Puri, Raul and LeGresley, Patrick and Casper, Jared and Catanzaro, Bryan},
  journal={arXiv preprint arXiv:1909.08053},
  year={2019}
}

@InProceedings{smoothquant,
    title = {{S}mooth{Q}uant: Accurate and Efficient Post-Training Quantization for Large Language Models},
    author = {Xiao, Guangxuan and Lin, Ji and Seznec, Mickael and Wu, Hao and Demouth, Julien and Han, Song},
    booktitle = {Proceedings of the 40th International Conference on Machine Learning},
    year = {2023}
}

@inproceedings{awq,
  title={AWQ: Activation-aware Weight Quantization for LLM Compression and Acceleration},
  author={Lin, Ji and Tang, Jiaming and Tang, Haotian and Yang, Shang and Chen, Wei-Ming and Wang, Wei-Chen and Xiao, Guangxuan and Dang, Xingyu and Gan, Chuang and Han, Song},
  booktitle={MLSys},
  year={2024}
}

@article{movementpa,
  title={Movement Pruning: Adaptive Sparsity by Fine-Tuning},
  author={Victor Sanh and Thomas Wolf and Alexander M. Rush},
  journal={ArXiv},
  year={2020},
  volume={abs/2005.07683},
  url={https://api.semanticscholar.org/CorpusID:218665313}
}

@ARTICLE{chang2021edgeaisurvey,
  author={Chang, Zhuoqing and Liu, Shubo and Xiong, Xingxing and Cai, Zhaohui and Tu, Guoqing},
  journal={IEEE Internet of Things Journal}, 
  title={A Survey of Recent Advances in Edge-Computing-Powered Artificial Intelligence of Things}, 
  year={2021},
  volume={8},
  number={18},
  pages={13849-13875},
  doi={10.1109/JIOT.2021.3088875}}

@article{gill2025edgeaisurvey,
  title={Edge AI: A taxonomy, systematic review and future directions},
  author={Gill, Sukhpal Singh and Golec, Muhammed and Hu, Jianmin and Xu, Minxian and Du, Junhui and Wu, Huaming and Walia, Guneet Kaur and Murugesan, Subramaniam Subramanian and Ali, Babar and Kumar, Mohit and others},
  journal={Cluster Computing},
  volume={28},
  number={1},
  pages={1--53},
  year={2025},
  publisher={Springer}
}

@article{llama2,
  title={Llama 2: Open foundation and fine-tuned chat models},
  author={Touvron, Hugo and Martin, Louis and Stone, Kevin and Albert, Peter and Almahairi, Amjad and Babaei, Yasmine and Bashlykov, Nikolay and Batra, Soumya and Bhargava, Prajjwal and Bhosale, Shruti and others},
  journal={arXiv preprint arXiv:2307.09288},
  year={2023}
}

@article{qwen2,
  title={Qwen2 Technical Report},
  author={An Yang and Baosong Yang and Binyuan Hui and Bo Zheng and Bowen Yu and Chang Zhou and Chengpeng Li and Chengyuan Li and Dayiheng Liu and Fei Huang and Guanting Dong and Haoran Wei and Huan Lin and Jialong Tang and Jialin Wang and Jian Yang and Jianhong Tu and Jianwei Zhang and Jianxin Ma and Jin Xu and Jingren Zhou and Jinze Bai and Jinzheng He and Junyang Lin and Kai Dang and Keming Lu and Ke-Yang Chen and Kexin Yang and Mei Li and Min Xue and Na Ni and Pei Zhang and Peng Wang and Ru Peng and Rui Men and Ruize Gao and Runji Lin and Shijie Wang and Shuai Bai and Sinan Tan and Tianhang Zhu and Tianhao Li and Tianyu Liu and Wenbin Ge and Xiaodong Deng and Xiaohuan Zhou and Xingzhang Ren and Xinyu Zhang and Xipin Wei and Xuancheng Ren and Yang Fan and Yang Yao and Yichang Zhang and Yunyang Wan and Yunfei Chu and Zeyu Cui and Zhenru Zhang and Zhi-Wei Fan},
  journal={ArXiv},
  year={2024},
  volume={abs/2407.10671},
  url={https://api.semanticscholar.org/CorpusID:271212307}
}

@inproceedings{first_spec,
  title={Fast inference from transformers via speculative decoding},
  author={Leviathan, Yaniv and Kalman, Matan and Matias, Yossi},
  booktitle={International Conference on Machine Learning},
  pages={19274--19286},
  year={2023},
  organization={PMLR}
}

@inproceedings{specinfer,
author = {Miao, Xupeng and Oliaro, Gabriele and Zhang, Zhihao and Cheng, Xinhao and Wang, Zeyu and Zhang, Zhengxin and Wong, Rae Ying Yee and Zhu, Alan and Yang, Lijie and Shi, Xiaoxiang and Shi, Chunan and Chen, Zhuoming and Arfeen, Daiyaan and Abhyankar, Reyna and Jia, Zhihao},
title = {SpecInfer: Accelerating Large Language Model Serving with Tree-based Speculative Inference and Verification},
year = {2024},
isbn = {9798400703867},
publisher = {Association for Computing Machinery},
address = {New York, NY, USA},
booktitle = {Proceedings of the 29th ACM International Conference on Architectural Support for Programming Languages and Operating Systems, Volume 3},
pages = {932–949},
numpages = {18},
location = {La Jolla, CA, USA},
series = {ASPLOS '24}
}

@inproceedings{eagle, 
    author = {Yuhui Li and Fangyun Wei and Chao Zhang and Hongyang Zhang}, 
    title = {{EAGLE}: Speculative Sampling Requires Rethinking Feature Uncertainty}, 
    booktitle = {International Conference on Machine Learning},
    year = {2024}
}

@inproceedings{eagle2, 
    author = {Yuhui Li and Fangyun Wei and Chao Zhang and Hongyang Zhang}, 
    title = {{EAGLE-2}: Faster Inference of Language Models with Dynamic Draft Trees}, 
    booktitle = {Empirical Methods in Natural Language Processing},
    year = {2024}
}

@inproceedings{medusa,
author = {Cai, Tianle and Li, Yuhong and Geng, Zhengyang and Peng, Hongwu and Lee, Jason D. and Chen, Deming and Dao, Tri},
title = {MEDUSA: Simple LLM inference acceleration framework with multiple decoding heads},
year = {2024},
publisher = {JMLR.org},
booktitle = {Proceedings of the 41st International Conference on Machine Learning},
articleno = {203},
numpages = {27},
location = {Vienna, Austria},
series = {ICML'24}
}

@INPROCEEDINGS{voltage,
  author={Hu, Chenghao and Li, Baochun},
  booktitle={2024 IEEE 44th International Conference on Distributed Computing Systems (ICDCS)}, 
  title={When the Edge Meets Transformers: Distributed Inference with Transformer Models}, 
  year={2024},
  volume={},
  number={},
  pages={82-92},
  doi={10.1109/ICDCS60910.2024.00017}}

@article{opt-tree,
  author       = {Jikai Wang and
                  Yi Su and
                  Juntao Li and
                  Qingrong Xia and
                  Zi Ye and
                  Xinyu Duan and
                  Zhefeng Wang and
                  Min Zhang},
  title        = {OPT-Tree: Speculative Decoding with Adaptive Draft Tree Structure},
  journal      = {Trans. Assoc. Comput. Linguistics},
  volume       = {13},
  pages        = {188--199},
  year         = {2025},
  url          = {https://doi.org/10.1162/tacl\_a\_00735},
  doi          = {10.1162/TACL\_A\_00735},
  bibsource    = {dblp computer science bibliography, https://dblp.org}
}

@ARTICLE{edgeshard,
  author={Zhang, Mingjin and Shen, Xiaoming and Cao, Jiannong and Cui, Zeyang and Jiang, Shan},
  journal={IEEE Internet of Things Journal}, 
  title={EdgeShard: Efficient LLM Inference via Collaborative Edge Computing}, 
  year={2024},
  volume={},
  number={},
  pages={1-1},
  doi={10.1109/JIOT.2024.3524255}}

@INPROCEEDINGS{galaxy,
  author={Ye, Shengyuan and Du, Jiangsu and Zeng, Liekang and Ou, Wenzhong and Chu, Xiaowen and Lu, Yutong and Chen, Xu},
  booktitle={IEEE INFOCOM 2024 - IEEE Conference on Computer Communications}, 
  title={Galaxy: A Resource-Efficient Collaborative Edge AI System for In-situ Transformer Inference}, 
  year={2024},
  volume={},
  number={},
  pages={1001-1010},
}

@INPROCEEDINGS{pipeinfer,
  author={Butler, Branden and Yu, Sixing and Mazaheri, Arya and Jannesari, Ali},
  booktitle={SC24: International Conference for High Performance Computing, Networking, Storage and Analysis}, 
  title={PipeInfer: Accelerating LLM Inference using Asynchronous Pipelined Speculation}, 
  year={2024},
  volume={},
  number={},
  pages={1-19},
  doi={10.1109/SC41406.2024.00046}}

@article{jupiter,
  title={Jupiter: Fast and resource-efficient collaborative inference of generative llms on edge devices},
  author={Ye, Shengyuan and Ouyang, Bei and Zeng, Liekang and Qian, Tianyi and Chu, Xiaowen and Tang, Jian and Chen, Xu},
  journal={arXiv preprint arXiv:2504.08242},
  year={2025}
}

@article{pipedec,
  title={PipeDec: Low-Latency Pipeline-based Inference with Dynamic Speculative Decoding towards Large-scale Models},
  author={Haofei Yin and Mengbai Xiao and Rouzhou Lu and Xiao Zhang and Dongxiao Yu and Guanghui Zhang},
  journal={arXiv preprint arXiv:2504.04104},
  year={2025}
}

@article{mtbench,
  title={Judging LLM-as-a-judge with MT-Bench and Chatbot Arena},
  author={Lianmin Zheng and Wei-Lin Chiang and Ying Sheng and Siyuan Zhuang and Zhanghao Wu and Yonghao Zhuang and Zi Lin and Zhuohan Li and Dacheng Li and Eric P. Xing and Haotong Zhang and Joseph E. Gonzalez and Ion Stoica},
  journal={ArXiv},
  year={2023},
  volume={abs/2306.05685},
  url={https://api.semanticscholar.org/CorpusID:259129398}
}

@article{humaneval,
  title={Evaluating Large Language Models Trained on Code},
  author={Mark Chen and Jerry Tworek and Heewoo Jun and Qiming Yuan and Henrique Pond{\'e} and Jared Kaplan and Harrison Edwards and Yura Burda and Nicholas Joseph and Greg Brockman and Alex Ray and Raul Puri and Gretchen Krueger and Michael Petrov and Heidy Khlaaf and Girish Sastry and Pamela Mishkin and Brooke Chan and Scott Gray and Nick Ryder and Mikhail Pavlov and Alethea Power and Lukasz Kaiser and Mo Bavarian and Clemens Winter and Philippe Tillet and Felipe Petroski Such and David W. Cummings and Matthias Plappert and Fotios Chantzis and Elizabeth Barnes and Ariel Herbert-Voss and William H. Guss and Alex Nichol and Igor Babuschkin and Suchir Balaji and Shantanu Jain and Andrew Carr and Jan Leike and Joshua Achiam and Vedant Misra and Evan Morikawa and Alec Radford and Matthew M. Knight and Miles Brundage and Mira Murati and Katie Mayer and Peter Welinder and Bob McGrew and Dario Amodei and Sam McCandlish and Ilya Sutskever and Wojciech Zaremba},
  journal={ArXiv},
  year={2021},
  volume={abs/2107.03374},
  url={https://api.semanticscholar.org/CorpusID:235755472}
}

@article{gsm8k,
  title={Training Verifiers to Solve Math Word Problems},
  author={Karl Cobbe and Vineet Kosaraju and Mo Bavarian and Mark Chen and Heewoo Jun and Lukasz Kaiser and Matthias Plappert and Jerry Tworek and Jacob Hilton and Reiichiro Nakano and Christopher Hesse and John Schulman},
  journal={ArXiv},
  year={2021},
  volume={abs/2110.14168},
  url={https://api.semanticscholar.org/CorpusID:239998651}
}

@misc{alpaca,
  author = {Rohan Taori and Ishaan Gulrajani and Tianyi Zhang and Yann Dubois and Xuechen Li and Carlos Guestrin and Percy Liang and Tatsunori B. Hashimoto },
  title = {Stanford Alpaca: An Instruction-following LLaMA model},
  year = {2023},
  publisher = {GitHub},
  journal = {GitHub repository},
  howpublished = {\url{https://github.com/tatsu-lab/stanford_alpaca}},
}

@inproceedings{cnn/dailtmail,
  title={Abstractive Text Summarization using Sequence-to-sequence RNNs and Beyond},
  author={Ramesh Nallapati and Bowen Zhou and C{\'i}cero Nogueira dos Santos and Çaglar G{\"u}lçehre and Bing Xiang},
  booktitle={Conference on Computational Natural Language Learning},
  year={2016},
  url={https://api.semanticscholar.org/CorpusID:8928715}
}

@article{natrual,title	= {Natural Questions: a Benchmark for Question Answering Research},author	= {Tom Kwiatkowski and Jennimaria Palomaki and Olivia Redfield and Michael Collins and Ankur Parikh and Chris Alberti and Danielle Epstein and Illia Polosukhin and Matthew Kelcey and Jacob Devlin and Kenton Lee and Kristina N. Toutanova and Llion Jones and Ming-Wei Chang and Andrew Dai and Jakob Uszkoreit and Quoc Le and Slav Petrov},year	= {2019},journal	= {Transactions of the Association of Computational Linguistics}}

@incollection{torch,
    title = {PyTorch: An Imperative Style, High-Performance Deep Learning Library},
    author = {Paszke, Adam and Gross, Sam and Massa, Francisco and Lerer, Adam and Bradbury, James and Chanan, Gregory and Killeen, Trevor and Lin, Zeming and Gimelshein, Natalia and Antiga, Luca and Desmaison, Alban and Kopf, Andreas and Yang, Edward and DeVito, Zachary and Raison, Martin and Tejani, Alykhan and Chilamkurthy, Sasank and Steiner, Benoit and Fang, Lu and Bai, Junjie and Chintala, Soumith},
    booktitle = {Advances in Neural Information Processing Systems 32},
    pages = {8024--8035},
    year = {2019},
    publisher = {Curran Associates, Inc.},
    url = {http://papers.neurips.cc/paper/9015-pytorch-an-imperative-style-high-performance-deep-learning-library.pdf}
}

@article{touvron2023llama,
  title={Llama 2: Open foundation and fine-tuned chat models},
  author={Touvron, Hugo and Martin, Louis and Stone, Kevin and Albert, Peter and Almahairi, Amjad and Babaei, Yasmine and Bashlykov, Nikolay and Batra, Soumya and Bhargava, Prajjwal and Bhosale, Shruti and others},
  journal={arXiv preprint arXiv:2307.09288},
  year={2023}
}
